\documentclass[a4paper,11pt]{article}  
\usepackage[toc,title]{appendix}
\pdfoutput=1

\usepackage{graphicx}  
\usepackage{bbold}
\usepackage{mathtools}
\usepackage{amsmath}
\numberwithin{equation}{section} 
\usepackage{amsfonts}
\usepackage{amssymb} 
\usepackage{slashed} 
\usepackage{color} 
\usepackage{xcolor}
\usepackage{tabularx}
\usepackage{hyperref}
\usepackage{bm}
\usepackage{cite}  
\usepackage{mathrsfs}
\usepackage{framed}

\def\eq#1{{Eq.~(\ref{#1})}}
\def\eqs#1#2{{Eqs.~(\ref{#1})--(\ref{#2})}}

\def\abs#1{\left| #1\right|}

\def\di{\mbox{d}}

\colorlet{grayline}{gray!70}
\definecolor{blueline}{rgb}{0,0.27,0.55}
\definecolor{DarkGray}{gray}{0.4}
\definecolor{Gray}{gray}{0.6}
\definecolor{oucrimsonred}{rgb}{0.6, 0.0, 0.0}
\definecolor{persianblue}{rgb}{0.11, 0.22, 0.73}
\definecolor{forestgreen}{rgb}{0.13,0.35,0.13}
 \hypersetup{colorlinks, citecolor=forestgreen, linkcolor=forestgreen, urlcolor=forestgreen}
\newcommand{\be}{\begin{equation}}
\newcommand{\ee}{\end{equation}}
\newcommand{\bea}{\begin{eqnarray}}
\newcommand{\eea}{\end{eqnarray}}

\newcommand*\xbar[1]{%
  \hbox{\;%
    \vbox{%
      \hrule height 0.5pt 
      \kern0.5ex
      \hbox{%
        \kern-0.25em
        \ensuremath{#1}%
        \kern-0.07em
      }%
    }%
  }%
} 
\newcommand{\com}[1]{}
\newcommand{\gsim}{\lower.7ex\hbox{$\;\stackrel{\textstyle>}{\sim}\;$}}
\newcommand{\lsim}{\lower.7ex\hbox{$\;\stackrel{\textstyle<}{\sim}\;$}} 

\newcommand{\bc}{\begin{center}}
\newcommand{\ec}{\end{center}}

\newcommand{\mWW}{m_{\scriptscriptstyle{W\!W}}}

\newcommand{\SM}{{\rm SM}}
\newcommand{\MW}{M_W}
\newcommand{\MZ}{M_Z}
\newcommand{\WW}{\scriptscriptstyle{W\!W}}

\newcommand{\sW}{\sin \theta_W}

\newcommand{\ssW}{s_{\W}}

\newcommand{\U}{\scriptscriptstyle{U}}
\newcommand{\D}{\scriptscriptstyle{D}}
\newcommand{\Ct}{c_{\Theta}}
\newcommand{\St}{s_{\Theta}}

\newcommand{\W}{\scriptscriptstyle{W}}

\newcommand{\betaW}{\beta_{\W}}

\newcommand{\CKM}{\scriptscriptstyle{\rm CKM}}
\newcommand{\LHC}{\scriptscriptstyle{\rm LHC}}
\newcommand{\HL}{\scriptscriptstyle{\rm HL}}
\newcommand{\TT}{\scriptscriptstyle{TT}}
\newcommand{\LL}{\scriptscriptstyle{LL}}

\begin{document}
\thispagestyle{empty}
\begin{center}
{\color{oucrimsonred}\Large {\bf 
    Testing the CKM unitarity at high energy via the $W^+W^-$ production

    \vspace{0.3cm}
    at the LHC  and future colliders}}

\vspace*{1.5cm}
{
{\bf E. Gabrielli$^{{a,b,c}}$}, 
 {\bf L. Marzola$^{{c,d}}$} and
{\bf K. M\"u\"ursepp$^{{c}}$} 
}\\

\vspace{0.5cm}
       {\small\it
(a) Physics Department, University of Trieste, Strada Costiera 11, \\ I-34151 Trieste, Italy
  \\[1mm]  
(b) INFN, Sezione di Trieste, Via Valerio 2, I-34127 Trieste, Italy
\\[1mm]   
(c) Laboratory of High-Energy and Computational Physics, NICPB, R\"avala pst 10, \\ 10143 Tallinn, Estonia
\\[1mm]
(d) Institute of Computer Science, University of Tartu, 
Narva mnt 18, 51009 Tartu, Estonia}

\ec

 \vskip0.5cm
\bc
{\color{DarkGray}
\rule{0.7\textwidth}{0.5pt}}
\ec
\vskip1cm
\bc
{\bf ABSTRACT}
\ec
We propose a novel test to assess the unitarity of the Cabibbo-Kobayashi-Maskawa matrix, $V_{\CKM}$, at present and future collider experiments. Our strategy makes use of the $W^+W^-$ production cross section to directly probe the $V_{\CKM}^\dagger V_{\CKM}$ product, which regulates the high-energy behavior of the observable. The violation of unitarity is signalled by an anomalous behavior of the cross section that grows quadratically with the $W^+W^-$ invariant mass with respect to the Standard Model prediction. By using the recent ATLAS measurements of the $W^+W^-$ cross section we are able to constrain the maximal unitarity violation allowed by current data, producing a bound complementary to the results of flavor physics experiments. Forecasts for the high luminosity phase of the LHC and for the future 100 TeV hadron collider are also discussed.

\vspace*{5mm}

\noindent

\newpage
\pagestyle{plain}

\newpage
\section{Introduction\label{sec:intro}} 
The Standard Model (SM) quark flavor mixing arises from a misalignment between the Yukawa coupling matrices of up- and down-type quarks. The effect is measurable in the charged-current (cc) interactions mediated by the $W^{\pm}$ gauge bosons and is parameterized by the Cabibbo-Kobayashi-Maskawa~\cite{Cabibbo:1963yz,Kobayashi:1973fv} (CKM) matrix, $V_{\CKM}$, appearing in the related interaction Lagrangian 
\be
   {\cal L}^{cc}=-\frac{g}{\sqrt{2}}\left( \bar{u}_L, \bar{c}_L,  \bar{t}_L\right)
     \gamma^{\mu} W_{\mu}^+ V_{\CKM}
\left(
\begin{array}{cc}
d_L & \\ s_L &  \\ b_L 
\end{array} \!\! \!\! \!\! \right)\, + h.c.\, ,\quad 
V_{\CKM}=\left(
\begin{array}{ccc}
V_{ud}~~ V_{us}~~ V_{ub} & \\
V_{cd}~~ V_{cs}~~ V_{cb} & \\
V_{td}~~ V_{ts}~~ V_{tb} 
\end{array}
 \!\! \!\! \!\!
\right)\, ,
\ee
where $u,c,t$ and $b,s,t$ stand for the corresponding up and down quark fields.

The CKM matrix is necessarily unitary within the SM: it is given by the product of the unitary matrices that connect gauge and mass eigenstates. The unitarity condition is expressed as 
\bea
\sum_i V_{ij} V^{\star}_{ik}=\delta_{jk}\quad\text{or}\quad  
\sum_j V_{ij} V^{\star}_{kj} = \delta_{ik}\, ,
\eea
where the indices $i$ and $j$ run on the up- or the down-type quark fields, respectively. Each of these conditions yields nine independent constraints, six of these can be represented as triangles in a complex plane. 

Experimental tests of unitarity in quark mixing probe combinations of the above conditions and have the potential to unveil the presence of new physics should anomalies be detected~\cite{Charles:2015gya}. For instance, models including new fields that mix with quarks in charged-current weak interactions typically entail a violation of unitarity at energies below the mass scale characteristic of the new degrees of freedom. Currently, independent measurements of the CKM matrix elements give for the 1st and 2nd row of the matrix~\cite{Workman:2022ynf},
\bea
\left|V_{ud}\right|^2+\left|V_{us}\right|^2+\left|V_{ub}\right|^2&=&0.9985 
\pm 0.0007 \label{CKMuu}\,,\\
\left|V_{cd}\right|^2+\left|V_{cs}\right|^2+\left|V_{cb}\right|^2&=&1.001\pm 0.012\label{CKMcc}\,,
\eea
while for the 1st and 2nd column we have~\cite{Workman:2022ynf},
\bea
\left|V_{ud}\right|^2+\left|V_{cd}\right|^2+\left|V_{td}\right|^2&=&0.9972
\pm 0.0020 \label{CKMdd}\\
\left|V_{us}\right|^2+\left|V_{cs}\right|^2+\left|V_{ts}\right|^2&=&1.004\pm 0.012 \, .
\label{CKMss}
\eea
Another independent test for the second row of the matrix, obtained from the measurement of $\sum_{u,c,d,s,b}\abs{V_{ij}}^2$ by subtracting the total contribution of the first row, gives~\cite{Workman:2022ynf}
\be
\left|V_{cd}\right|^2+\left|V_{cs}\right|^2+\left|V_{cb}\right|^2=1.002\pm 0.027\label{CKMcc2}\, .
\ee

In order to quantify possible deviations from unitarity in charged current interactions, we introduce the diagonal parameters $\delta_q$, with $q=u,d,s,c,b$, defined as
\bea
\label{deltau}
\delta_{q} \equiv
\begin{cases}
  \left|V_{qd}\right|^2+\left|V_{qs}\right|^2+\left|V_{qb}
  \right|^2-1 & {\rm for}~~q=u,c
  \\\\
  \left|V_{uq}\right|^2+\left|V_{cq}\right|^2+\left|V_{tq}
  \right|^2-1&{\rm for}~~q=d,s,b\, .  
\end{cases}
\eea
As we can see from the above results, the CKM unitarity tests are presently well in agreement with SM expectations, although a $2.2\sigma$ tension concerning $\delta_u$ is present. The discrepancy, dubbed ``Cabibbo angle anomaly",
\cite{Hardy:2020qwl,Grossman:2019bzp,Kirk:2020wdk,Manzari:2021kma,Crivellin:2022ctt,Cirigliano:2022yyo,Crivellin:2022rhw} is mainly due to the recent determination of the $\left|V_{ud}\right|$ element from the super-allowed $0^+ \to 0^+$ nuclear $\beta$ decays~\cite{Hardy:2020qwl}.  However, it is fair to say that the discrepancy depends on which $V_{ud}$ and $V_{us}$  measurements are considered in global fit. For instance, an updated analysis in Ref.~\cite{Crivellin:2022rhw,Cirigliano:2022yyo} finds $\delta_{u}= -0.00151 (53)$, corresponding to a $2.85\sigma$ deviation that could be further amplified by using most constraining measurements of $V_{ud}$ and $V_{us}$ in the global fit. We refer the reader to Refs.~\cite{Cirigliano:2023nol,Kitahara:2023xab,Alves:2023ufm,Belfatto:2023tbv,Crivellin:2022rhw,Seng:2021gmh,Alok:2021ydy,Chang:2021axw,Crivellin:2021rbf,Branco:2021vhs,Belfatto:2021jhf,Crivellin:2021njn,Crivellin:2020lzu,Capdevila:2020rrl,Coutinho:2019aiy,Belfatto:2019swo} for the various new physics interpretations of the Cabibbo angle anomaly.

Currently, the CKM matrix unitarity tests encoded in the $\delta_{q}$ parameters rely on the cancellation between the unity and the sum of the involved squared matrix elements, which are generally individually measured. This methodology affects the power of the test as the outcome strongly depends on the precision level at which each matrix element is measured. It would then be desirable to have a further independent test of unitarity that could directly gauge the parameters $\delta_q$, thereby bypassing the problems posed by the determination of the individual contributions.  As we show in the following, measurements of the $W^+W^-$ production cross section at hadron colliders provide the means to achieve exactly that.

The process $p p \to W^+W^-$ is known to be a powerful tool for constraining new physics models affecting SM precision observables~\cite{ATLAS:2012mec,ATLAS:2016zwm,CMS:2015tmu,CMS:2013ant,Degrande:2012wf,Franceschini:2017xkh}. In order to test unitarity in charged current interactions, we make use of the related tree-level amplitude and analyze the interplay between the divergent $s$-channel term, driven by the longitudinal modes of the produced gauge bosons, and the quark contributions weighted by the CKM matrix elements under the customary assumption that neutral currents are not modified. Whereas gauge invariance and the unitarity of the CKM matrix ensure an exact cancellation in the SM, the cancellation  does not occur in the presence of non-vanishing $\delta_q$ parameters. As a result, the loss of unitarity yields an anomalous contribution to the cross section that grows quadratically with the partonic center of mass energy or, analogously, with the $W^+W^-$ invariant mass. Importantly, the magnitude of the anomalous term is proportional to $\delta_q$ and measurements of the $p p \to W^+W^-$ cross section at colliders can then be used to directly constrain the parameter. 

Recent measurements of this observable by the CMS
\cite{CMS:2021pqj,CMS:2013piy,CMS:2020mxy} and ATLAS~\cite{ATLAS:2014xea,ATLAS:2017bbg,ATLAS:2019rob,ATLAS:2023zis} collaborations are found to be in good  agreement with SM predictions~\cite{Gehrmann:2014fva,Grazzini:2016ctr,deFlorian:2016uhr,Grazzini:2019jkl}. We utilize the data to assess the power of the LHC to constrain unitarity in the quark mixing and extend the analysis to gauge the reach of the planned high-luminosity phase, as well as of the future hadron collider running at $100~{\rm TeV}$~\cite{Mangano:2017tke,FCC:2018byv}.

We remark that our procedure for constraining the  $\delta_q$ parameters encoding the CKM unitarity violation assumes the presence of only one non-vanishing contribution at a time, on top of disregarding other potential sources of unitarity violation that could stem, for instance, from an anomalous triple gauge boson coupling.

\section{WW production via quark fusion 
\label{sec:DY}}  

The $W^+W^-$ states of interest are mainly produced at proton colliders via electroweak processes in a continuous range of diboson invariant masses, $\mWW$. The corresponding hadronic differential cross section can be written as
\be
\frac{\di^2 \sigma}{\di \Omega \,\di \mWW} =
\frac{\betaW}{64 \pi^2 \, \mWW^2}
  \sum_{q=u,d,s,c,b} |\overline{\cal M}^{\,q \bar q}|^2\, L^{q\bar q}(\tau) \, ,  
\label{diffcross}
\ee
where $\betaW=\sqrt{1-4\MW^2/\mWW^2}$, $\tau=\mWW/\sqrt{S}$, $\MW$ is the $W$ boson mass and $\sqrt{S}$ is the center of mass energy of the $pp$  system. The infinitesimal solid angle expressed in polar coordinates is $\di\Omega=\di\cos\Theta~\di\varphi$, with $\Theta$ being the scattering angle in the center of mass frame of reference (CM) for the partonic process $q\bar{q}\to W^+W^-$ and $\varphi$ the azimuthal angle. The unpolarized squared amplitude $|\xbar{\mathcal{M}}^{\,q \bar q}|^2$ of the partonic process, mediated over initial states, is weighted by the parton luminosity $L^{q\bar q}(\tau)$ of the specific $q\bar{q}$ initial state  
\be
L^{q\bar q} (\tau)= \frac{4 \tau}{\sqrt{S}} \int\limits_{\tau}^{1/\tau} \frac{\di z}{z} \, q_{q} (\tau \,z) \, q_{\bar q} \left( \frac{\tau}{z}\right)\, ,
\label{Lumi}
\ee
where $q_q(x)$ and  $q_{\bar q}(x)$ are the parton distribution function (PDF) of the parton $q$ and antiparton $\bar{q}$, respectively. We use the recent {(\tt PDF4LHC21)} release~\cite{PDF4LHCWorkingGroup:2022cjn}, for $\sqrt{S}=13$ TeV and factorization scale $\mWW$, when deriving the unitarity bounds with current LHC data. 

\begin{figure}[h!]
  \begin{center}
  \includegraphics[width=6.0in]{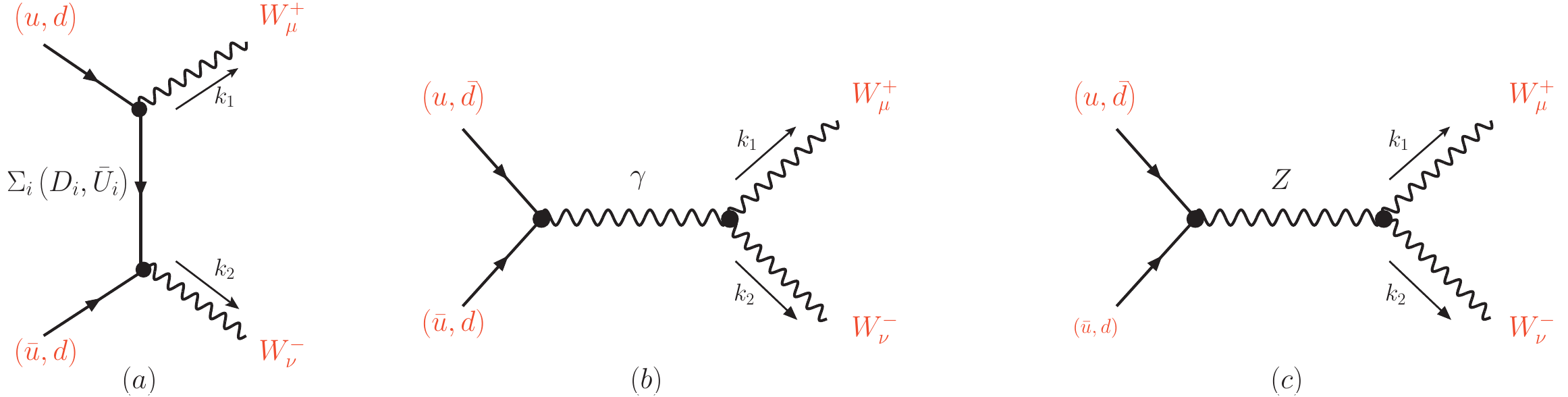}
  \caption{\small Feynman diagrams for the processes $p\; p\to W^+W^-$ initiated by first generation quarks. The symbols $U_i=\{u,c,t\}$ and $D_i=\{d,s,b\}$ indicate up and down-type quarks, respectively. We neglect diagrams mediated by the Higgs boson considering the limit of massless quarks in the initial state. The arrows on the fermion lines indicate the momentum flow. We indicate with $k_1$ and $k_2$ the (outgoing) $W^+$ and $W^-$ 4-momenta.
  \label{fig:DYWW} 
  }
  \end{center}
  \end{figure}

The tree-level Feynman diagrams contributing to the underlying partonic process $\bar{q}(p_1) q(p_2) \to W^+(k_1) W^-(k_2)$ are shown in Fig.~\ref{fig:DYWW} for the case of first generation quarks. For the purpose of the present analysis we can work in the high energy approximation in which all quark masses can be neglected---including the top-quark mass, as discussed at the end of this section. The amplitude for the process is then written as 
\bea
{\cal M}^{q\bar q}&=&-ie^2 \Big[\bar{v}(p_1) \Gamma^q_{\mu\nu} u(p_2)\Big]
\varepsilon^{\mu}(k_1)^{\star}\varepsilon^{\nu}(k_2)^{\star}
\, ,
\label{MDYWW}
\eea
where we indicated with $\varepsilon^{\mu}(k_1)$ and $\varepsilon^{\nu}(k_2)$ the polarization vectors of $W^+$ and $W^-$, respectively. The effective vertex $\Gamma^q_{\alpha\beta}$ is given under the massless quark approximation by
\bea
\Gamma^{q=u,c}_{\mu\nu}&=&\frac{1}{s}
\left(\gamma^{\alpha} \bar{g}_V^q-\gamma^{\alpha}\gamma_5 \bar{g}_A^q\right)
V_{\alpha\nu\mu}(p,-k_2,-k_1)+\frac{\xi_{q}}{4 t \ssW^2}\gamma_{\nu}\left(\slashed{p}_2-\slashed{k}_1\right)\gamma_{\mu}(1-\gamma_5)\,,
\nonumber\\
\Gamma^{q=d,s,b}_{\mu\nu}&=&\frac{1}{s}
\left(\gamma^{\alpha} \bar{g}_V^q-\gamma^{\alpha}\gamma_5 \bar{g}_A^q\right)
V_{\alpha\nu\mu}(p,-k_2,-k_1)+\frac{\xi_{q}}{4 u \ssW^2}\gamma_{\nu}\left(\slashed{p}_2-\slashed{k}_2\right)\gamma_{\mu}(1-\gamma_5)\,,
\label{Gamma}
\eea
with $\ssW= \sW$ and $\theta_W$ the Weinberg angle, $e$ being the unit  electric charge and $p=k_1+k_2$. The parameter $\xi_q\equiv \sum_{\D=d,s,b} \left[ V^*_{q \D}  V_{q \D}\right]$ for $q=u,c$ and $\xi_q\equiv \sum_{\U=u,c,t} \left[ V^*_{\U q}  V_{\U q}\right]$ for $q=d,s,b$, model the CKM unitarity condition---$\xi_q=1$ in the SM---and arise from the diagram (a) in Fig.~\ref{fig:DYWW}. The effective couplings $\bar{g}^q_{V,A}$ in \eq{Gamma} are given by
\be
\bar{g}_V^q=Q^q+\frac{g_V^q\chi}{\ssW^2}\, ,~~
\bar{g}_A^q=\frac{g_A^q\chi}{\ssW^2}\, ,
~~ \chi=\frac{s}{2(s-\MZ^2)}\, ,
\label{effgva}
\ee
where  $\MZ$ is the $Z$ boson mass, $g_{V}^q=T_3^q-2Q^q\ssW^2$, $g_{A}^q=T_3^q$ and $T_3^q$ and $Q^q$ are the isospin and electric charge (in unit of $e$) of the quark $q$. The $\chi$ term in \eq{effgva}, which weights the contribution of the virtual $Z$ 
channel, is real since we neglect the $Z$ width contribution. The function $V_{\alpha\nu\mu}(k_1,k_2,k_3)$ is the Feynman rule of the trilinear vertex $V_{\alpha}(k_1)~W^+_{\nu}(k_2)~W^-_{\mu}(k_3)$, $V\in\{\gamma ,Z\}$, given by 
\be
V_{\alpha\nu\mu}(k_1,k_2,k_3)=(k_1-k_2)_{\mu}g_{\alpha\nu}
+(k_2-k_3)_{\alpha} g_{\mu \nu}+(k_3-k_1)_{\nu} g_{\alpha\mu}\, ,
\label{V3}
\ee
for incoming momenta in the trilinear vertex: $(k_1+k_2+k_3=0)$. The Mandelstam variables $s,t$ and $u$ appearing in the above equations are given by
\be
s=(p_1+p_2)^2, \quad t=(p_2-k_1)^2, \quad u=(p_2-k_2)^2, 
\ee
where $s$ coincides with the diboson invariant mass $s=\mWW^2$.

In order to investigate the effects of unitarity violation we introduce the parametrization 
\be
\xi_q=1+\delta_q
\label{xiq}
\ee
in the effective vertex $\Gamma^q_{\mu\nu}$ defined in \eq{Gamma} and for $\delta_q$ given in \eq{deltau}. Then, the unpolarized square amplitude for the process $q\,\bar{q}\to W^+W^-$, mediated over initial spin and color degrees of freedom, is given by
\be
|\xbar{{\cal M}}^{\; q\bar q}|^{2}(\delta_q)=|\xbar{{\cal M}}^{\; q\bar q}_{\SM}|^{2}+
\delta_q ~\Delta \xbar{{\cal M}}^{\; q\bar q}_{1} +
\delta^2_q ~\Delta \xbar{{\cal M}}^{\; q\bar q}_{2}\,,
\label{M2tot}
\ee
where $|\xbar{{\cal M}}^{\; q\bar q}_{\SM}|^{2}$ is the SM contribution\footnote{The explicit expression can be found in Ref.~\cite{Workman:2022ynf}.} and  the extra terms vanish in absence of unitarity violation. By retaining only the leading orders in the $s/\MW^2 \gg 1$ expansion, we find
\bea
\Delta \xbar{{\cal M}}^{\; q\bar q}_{1}&\simeq&
 \frac{\alpha_{\W}^2\pi^2}{6 N_c }\left(\frac{s}{\MW^2} \right)
\St^2\left[6-\frac{\MZ^2}{\MW^2}\left(3-4x_q\ssW^2\right)\right]\, ,
\nonumber\\
   \Delta \xbar{{\cal M}}^{\; q\bar q}_{2}&\simeq&
   \frac{\alpha_{\W}^2\pi^2}{4N_c}
   \Big[
     \left(\frac{s^2}{\MW^4}\right)\St^2+
        \left(\frac{s}{\MW^2}\right)4\left(3+\Ct^2\right)\Big]\, ,
   \label{DeltaM}
   \eea
where $\alpha_{\W}=\alpha/\ssW^2$. In the equation above, $x_q=1,\,1/2$ for up- (U) and down-type (D) quarks respectively and $N_c=3$ is the number of colors. We also defined $\Ct=\cos{\Theta}$ and $\St=\sin{\Theta}$, with $\Theta$ being the scattering angle formed by the quark and $W^-$ momenta in the CM frame. In our numerical analysis we use the exact analytical results for the above anomalous contributions, given in the appendix~\ref{sec:Appendix} as functions of $s$ and $\Theta$.

Concerning the parametrization in \eq{xiq}, we perform our analysis considering constant values for the parameters $\delta_q$, that is independent of the partonic CM energy $\sqrt s$. This scaling is expected to hold at energies below the characteristic scale $\Lambda$ where the new physics effects restore unitarity. We assume $\Lambda$ to be much larger than the energies explored at the considered collider experiments and do not speculate on the origin of this effective scale to retain independence from the specifics of the UV complete theory. Prototypical new physics models that could induce the loss of unitarity in the CKM mixing include heavy fourth-generation particles or vector-like quarks that mix with the SM particles, for instance, via Yukawa interactions. Explicit examples are provided, for instance, by universal see-saw models~\cite{Davidson:1987mh,Davidson:1987tr,Davidson:1989bx} and their left-right symmetric extensions \cite{Berezhiani:1983hm,Dimopoulos:1983rz}.

As anticipated, we neglect the top-quark mass ($m_t$) corrections that appear in processes initiated by down-type quarks through a virtual top exchange. For $d\bar{d}$ and $s\bar{s}$ initial states, these contributions are proportional to ${\cal O}(m_t^2/\mWW^2 |V_{td}|^2)$ and ${\cal O}(m_t^2/\mWW^2 |V_{ts}|^2)$, respectively, in both the SM and the anomalous contributions. Since the relevant $\mWW$ region for constraining the $\delta_{d,s}$ parameters lies in the range of invariant masses $\mWW \gsim {\rm 1~TeV}$, taking into account the suppression of the off-diagonal terms in the CKM matrix elements yields corrections well below the per mille level. For $b\bar{b}$ quark initial states, instead, $V_{tb}\sim 1$ results in much larger top mass corrections of the order of a few percent in the relevant region $\mWW \gsim {\rm TeV}$. However, since the bounds provided with our method on $\delta_b$ are at a 5-10\% precision level, we can still safely neglect these top-quark mass effects.

In the following analysis we use the cross section in \eq{diffcross} integrated on the whole solid angle $\Omega$. However, suitable angular cuts could in principle be used to increase the sensitivity of the cross section to the CKM unitarity violating terms $\delta_q$. This is demonstrated in Fig.~\ref{fig:angulardist}, where we plot with a solid line the SM contribution to the angular distribution \eqref{diffcross} as a function of $\cos{\theta}$ for $\mWW=1$ TeV. The pure $\delta_{u}$ contribution to the same differential cross section is indicated by the dashed lines given for the indicated benchmark values of the parameter. Each distribution is normalized so that its angular integral yields a unit value. The remaining $\delta_q$ parameters were set to vanishing values when producing the plot, but qualitatively similar distributions are obtained by considering each $\delta_q$ contribution in isolation. As we can see from these results, the new physics contribution peaks in the central region of the detector ($\theta=\pi/2$) where the SM contribution is minimum. Consequently, kinematic cuts apt to select the region $|\cos{\theta}| \lsim 0.8$ could enhance the sensitivity to the $\delta_q$ terms.

\begin{figure}[h!]
\begin{center}
  \vspace{-1.5cm}
  \includegraphics[width=.49\linewidth]{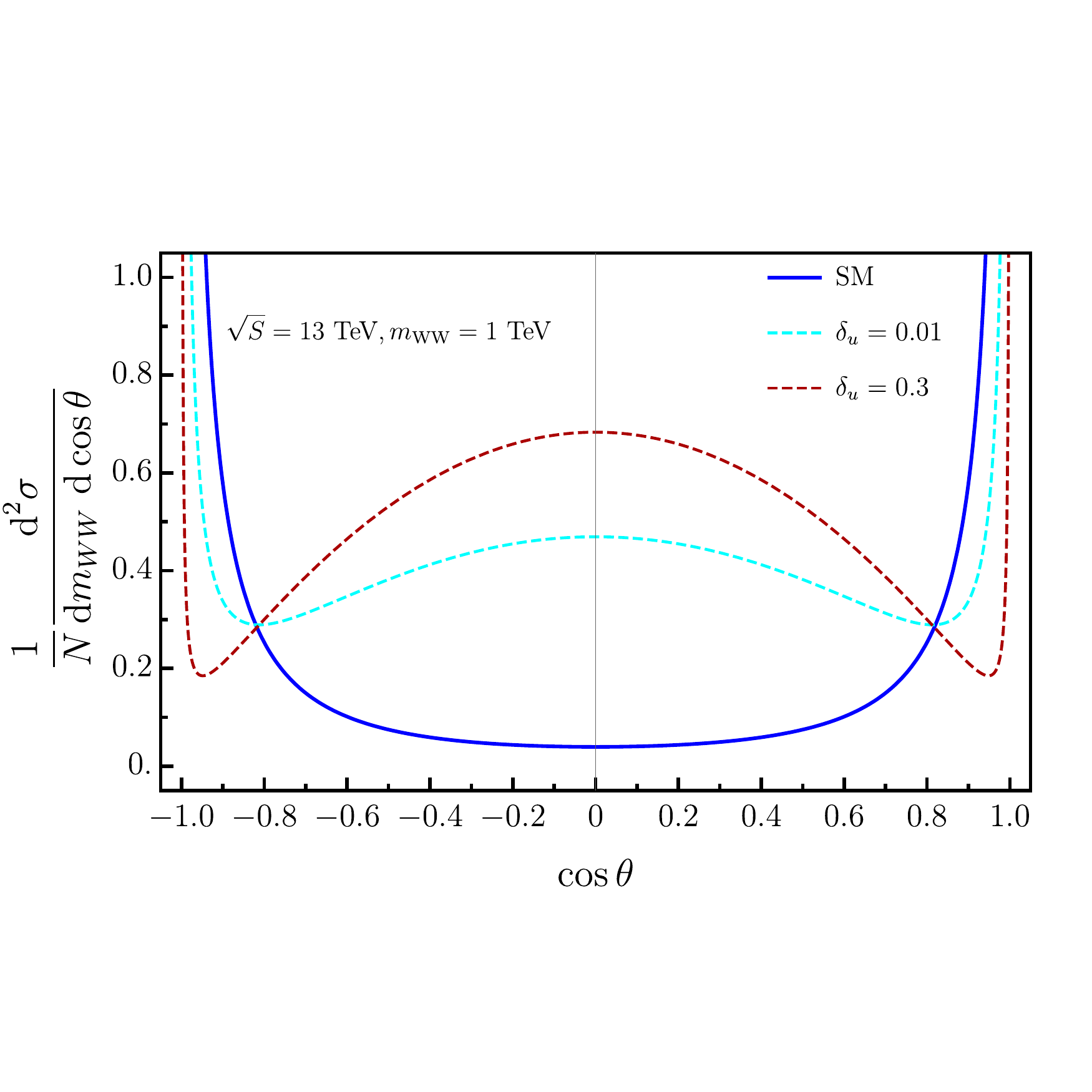}
  \vspace{-1.5cm}
  \caption{\small Normalized contributions to the the normalized differential cross section in \eq{diffcross} evaluated at $\mWW=1$ TeV. The blue solid line correspond to the SM case ($\delta_u=0$), while the dashed lines show pure $\delta_u$ contributions given for $\delta_u=0.01$ (light blue) and $\delta_u=0.3$ (red). Each contribution is individually normalized so that the corresponding angular integral yields a unit value.
\label{fig:angulardist} 
}
\end{center}
\end{figure}

\subsection{VBF background and theoretical uncertainties}
The $W^+W^-$ final states of interest must necessarily be produced via quark fusion, which is sensitive to the CKM contributions targeted by the test. A potentially large background for this analysis---especially important at the large diboson invariant masses that could be achieved in future collider experiments---is given by the vector boson fusion (VBF) proceeding via the exchange of two virtual $W$ or $Z$ bosons to give 
\be
p~p\to W^+W^- j_1~j_2\,,
\ee
where the two jets, $j_{1,2}$, are typically very forward and could fall outside of the acceptance of the detector thereby faking the signal of interest.  

To gauge the size of this background we resort to the Effective Vector Boson Approximation (EVBA). This computational scheme replaces the virtual bosons emitted from the initial quarks lines with real particles, emitted with probabilities that depend on their polarization, thereby discriminating in this case between the longitudinal and transverse contributions of the intermediate gauge bosons. We simplify our estimate by retaining only the contributions provided by the polarized process $W^+W^-\to W^+W^-$ for the longitudinal ($L$) and transverse ($T$) $W$ polarizations. The $ZZ\to WW$ contributions are sub-leading with respect to the $WW\to WW$ one and can therefore be neglected. The EVBA computation of the cross section is known to overestimate the actual value since no upper bounds on the transverse momenta of the two jets are imposed with this method. In this respect, the calculation provides a conservative estimate of the VBF background for the $q\bar{q}\to WW$ process of interest.

Using the results given in Ref.~\cite{Green:2003if}, we obtain for the intermediate polarized $W^{\pm}$ contribution to the $pp\to W^+W^-j_1j_2$ differential cross section 
\bea
\frac{d \sigma^{\rm VBF}}{d \mWW} &=& 
\frac{1}{2}~ \int_{M^2/S}^1\frac{d z}{z} ~  \sum_{q=u,d,s,c,b}
L^{q\bar{q}}(\sqrt{\hat{s}}) ~\sqrt{a}~
\Big\{
L^{\WW}_{q\bar{q}}(\hat{s})~ \sigma^{\LL}_{\WW}(\mWW^2)+
T^{\WW}_{q\bar{q}}(\hat{s})~ \sigma^{\TT}_{\WW}(\mWW^2)\Big\},~~~~~~~
\label{eq:EVBA}
\\
    L^{\WW}_{q\bar{q}}(\hat{s})&=&-\left(\frac{\alpha_{\W}}{4\pi }\right)^2\frac{1}{a}
    \left[(1+a)\ln{a}+2(1-a)\right]\,,
\nonumber\\
T^{\WW}_{q\bar{q}}(\hat{s})&=&-\left(\frac{\alpha_{\W}}{8\pi}\right)^2\frac{1}{a}
\left[\ln{\left(\frac{\hat{s}}{m^2_{\W}}\right)}\right]^2
    \left[(2+a)^2\ln{a}+2(1-a)(3+a)\right]\,,
\eea
where  $L^{WW}_{q\bar{q}}$ and $T^{WW}_{q\bar{q}}$ the effective luminosities for the $q\bar{q}$ system to emit longitudinal and transversely polarized $WW$ pairs \cite{BP}, respectively, $a=\mWW^2/\hat{s}$, $\hat{s}=z S$ is the initial $q\bar{q}$ invariant mass, $\sqrt{S}$ the $pp$ center of mass energy and $L^{q\bar{q}}(\sqrt{\hat{s}})$ is the parton luminosity for initial $q\bar{q}$ quark state as defined in \eq{Lumi}.  Finally, the total cross sections
$\sigma^{TT}_{\WW}$ and $\sigma^{LL}_{\WW}$, appearing in \eq{eq:EVBA}, are given by~\cite{Denner:1997kq}
\bea
\sigma^{\LL}_{\WW}(s)&=&
\frac{\pi\alpha_{\W}^2}{s}
\left\{
\frac{\left(M_H^2+M_Z^2\right)\left(2M_Z^2L_c+M_H^2 c\right)}{4M_W^2}
+\frac{c\left(75-26c^2-c^4\right)}{48 c_W^4(1-c^2)}
+\frac{c\left(3 + c^2\right)}{6}
\right\}\,,
\nonumber\\
\sigma^{\TT}_{\WW}(s)&=&
\frac{\pi\alpha_{\W}^2}{s}
\left\{
\frac{c\left(75 - 26 c^2 - c^4\right)}{3\left(1 - c^2\right)} 
 + \frac{c\left(3 + c^2\right)}{24} + 8L_c
\right\}\,,
\eea
with $c_W=\cos{\theta_W}$, $M_H$ being the Higgs boson mass and $L_c=\ln{\left[(1-c)/(1+c)\right]}$, where $c=\cos{\theta_{\rm cut}}$ and $\theta$ is the
$WW$ scattering angle in the $pp$ center of mass frame. 

In the computation of $\sigma^{AA}_{\WW}$, with $A=\{L,T\}$, we retained only the dominant contributions provided by same polarizations of $WW$ initial states given by $\sigma^{AA}_{\WW}=\sigma(W^+_AW^-_A\to W^+_LW^-_L) + \sigma(W^+_AW^-_A\to W^+_TW^-_T)$, where the corresponding polarized cross sections are provided in~\cite{Denner:1997kq}.
The contribution of intermediate $ZZ$ bosons can be computed in a similar fashion but, as mentioned, is made subdominant by the involved couplings with quarks.

Estimating the VBF contribution through the above method, we find that the yield of the process is a few percent of the quark fusion contribution at the LHC on the analyzed invariant diboson mass window. At the FCC-hh the relative yield of VBF with respect to the vector boson fusion reaches at most the 20\% level. However, this background can be reduced again to the percent level on the whole (extended) diboson mass range by imposing a cut to exclude events yielding $W$ bosons in cones with a 60 degree opening centered along the forward and backward beam directions. As such, we believe that the VBF contribution cannot pollute the signal enough to hinder the proposed analysis at the LHC nor at future colliders. 

As for the dominant sources of uncertainties in the $pp\to W^+W^-$ process, we refer the reader to the analysis of Ref.~\cite{Grazzini:2019jkl}, where the cross section has been computed at the NNLO QCD and NLO EW corrections level. As shown in the next section, in our computation of the cross section we included these corrections by rescaling the leading order (LO) cross section with a multiplicative $K$-factor. According to the analysis of \cite{Grazzini:2019jkl} (cfr. their table 3) the relative uncertainty in the total cross section, absorbed in our analysis in the $K$-factor, is of the 2-3 percent level. This uncertainty includes the ones affecting the relevant SM inputs, the missing higher order corrections and the uncertainty due to the scale variation in the quark PDF functions. We neglect the small variations in the $K$-factor over the analyzed $\mWW$ range highlighted by the results of Ref.~\cite{Grazzini:2019jkl} which includes the dominant NNLO QCD corrections (see Fig. 7 of \cite{Grazzini:2019jkl} for more details).

\vspace{0.5cm}
\section{Bounds on the unitarity violation parameters}
\label{sec:Results} 
The ATLAS~\cite{ATLAS:2019rob,ATLAS:2023zis} and CMS~\cite{CMS:2020mxy,CMS:2024cfq} collaborations have recently measured the $pp \to W^+W^-$ cross section at the LHC run2 using ${\cal L}\sim 140~{\rm fb}^{-1}$ and ${\cal L}\sim 36~{\rm fb}^{-1}$ integrated luminosities, respectively. Results for the total and differential cross sections are in good agreement with the corresponding SM predictions~\cite{Gehrmann:2014fva,Grazzini:2016ctr,deFlorian:2016uhr,Grazzini:2019jkl}. 

The expressions provided in section \ref{sec:DY} for the diboson cross section are obtained at the leading order (LO) in QCD. Higher order corrections can be included via the multiplicative $K$-factor relating the LO and NNLO cross section that, for $\sqrt{S}=13~{\rm TeV}$, is $K\simeq 1.721$~\cite{Grazzini:2019jkl}. We assume that the $K$ factors enhance the anomalous terms in the same way as the SM contribution.  The mild dependence of the NNLO corrections on $\mWW$~\cite{Gehrmann:2014fva,Grazzini:2016ctr,Grazzini:2019jkl} can be safely neglected on the considered range of the diboson invariant masses. 

\subsection{LHC: run2 and high-luminosity phase}

Following the ATLAS analyses in Refs.~\cite{ATLAS:2019rob,ATLAS:2023zis,ATLAS}, we consider six bins $\Delta_i$ for the diboson invariant mass, with edges given by
\be
\small
\Delta_i=\Big\{
\left[2\MW/{\rm GeV},~200\right],~~\left[200,~300\right],~~\left[300,~450\right],~~\left[450,~600\right],~~\left[600,~1200\right],~~\left[> 1200\right]
\Big\}\, ,
\label{LHCbins}
\ee
see Fig.14h in~\cite{ATLAS} (where $m_{T,e\mu}$ coincides with our $\mWW$ in the $W^+W^-$ CM frame). To constrain the unitarity violation parameters $\delta_q$ we then utilize a $\chi^2$ test set at the 95\% confidence level (CL)
\be
\chi^2(\delta_q)=\sum_{i=1}^6 \frac{\Big(\alpha_i\,\bar{\sigma}_i(\delta_q) -r_i\Big)^2}{\varepsilon_i^2}\leq \, 12.592\,, 
\label{chi2LHC}
\ee
where the index $i$ runs over the specified bins $\Delta_i$ with experimental error $\varepsilon_i$ and central values $r_i$ for the ratio of data/SM yield, reported in Tab.~\ref{tab:data}. The cross sections $\bar{\sigma}_i(\delta_q)$ are obtained by integrating \eq{diffcross} over the solid angle and in each bin: 
\be
\bar{\sigma}_i(\delta)\equiv \int_{\Delta_i} 
\frac{\di \sigma(\delta)}{\di \mWW \di\Omega} \, \di \mWW\,\di\Omega \,.
\label{sigmabin}
\ee
The value obtained for $\delta_q=0$ recovers the corresponding SM cross section $\bar{\sigma}^{\rm SM}_i\equiv \bar{\sigma}_i(0)$, which we utilize to compute the factors $\alpha_i\equiv r_i/\bar{\sigma}_{\rm SM}$ that correct for the normalization used in the experimental data, the collider luminosity as well as for the branching ratio and selection efficiencies involved in the search. 

We assume that the parameters $\delta_q$ are independent quantities and perform  $\chi^2$ tests by retaining only one non-vanishing contribution at a time. The  results obtained are reported in Fig. \ref{fig:UU-LHC}, in which we plot the $\chi^2$ value as a function of the considered $\delta_q$ parameter for the cases of $q=u,c,d,s$. The projections for the high-luminosity phase of the LHC (HL-LHC) are obtained upon a rescaling of the experimental errors by a factor of $\sqrt{{\cal L}_{\LHC}/{\cal L}_{\HL}}$, where we take ${\cal L}_{\LHC}=140~{\rm fb}^{-1}$ and ${\cal L}_{\HL}=3000~{\rm fb}^{-1}$. The bounds for all $\delta_q$ parameters  obtained at  95\% CL are summarized in Tab.~\ref{tab:delta-LHC}.

The up- and down-quark dominance in the proton PDF results in a larger sensitivity to the $\delta_u$ and $\delta_d$ parameters constrained below 2--4\%, with the run2 data, and up to 0.3--0.6\% with the increased luminosity of HL-LHC. Milder bounds are obtained for $\delta_s$ and $\delta_c$, of order $10\%$ and $20\%$ with the run2 data and up to $2\%$ and $5\%$ at the HL-LHC, respectively. As for $\delta_b$, looser constraints of about 30\% and 10\% are obtained in the two cases. The bounds in Tab.~\ref{tab:delta-LHC} are consistent with the results in \eqs{CKMuu}{CKMss} due to flavor physics, which still provides the strongest bounds on the $\delta_{u,d,s,c}$ parameters. We stress that the bounds in \eqs{CKMuu}{CKMss} are obtained through individual measurements of the involved $V_{\CKM}$ matrix elements. Conversely, the method suggested here directly probes the sum encapsulated in $\delta_q$ and, therefore, provides an independent complementary test that could be used along with the flavor physics results.

\begin{table}[h!]
  \centering
\begin{tabular}{ccc}
  $\Delta_i$  &   $r_i$ \hskip1cm & $\epsilon_i$ \hskip1cm \\[0.2cm]
\hline\\
$\left[2\MW/{\rm GeV},~200\right]$& 1.026 & 0.013\\
$\left[200,~300\right]$& 1.037 & 0.014\\
$\left[300,~450\right]$& 1.015 & 0.013\\
$\left[450,~600\right]$& 0.997 & 0.017\\
$\left[600,~1200\right]$& 1.013 & 0.023\\
$\left[> 1200\right]$& 0.994 & 0.120\\
   \hline%
\end{tabular}
\caption{\small\label{tab:data} The diboson invariant mass bins $\Delta_i$, corresponding central values $r_i$ of the data/SM yield ratios and related errors $\varepsilon_i$  used in the LHC run2 analysis, see Fig.14h in~\cite{ATLAS}.}
\end{table}

\begin{figure}[h!]
\begin{center}
  \vspace{-1.5cm}
  \includegraphics[width=.49\linewidth]{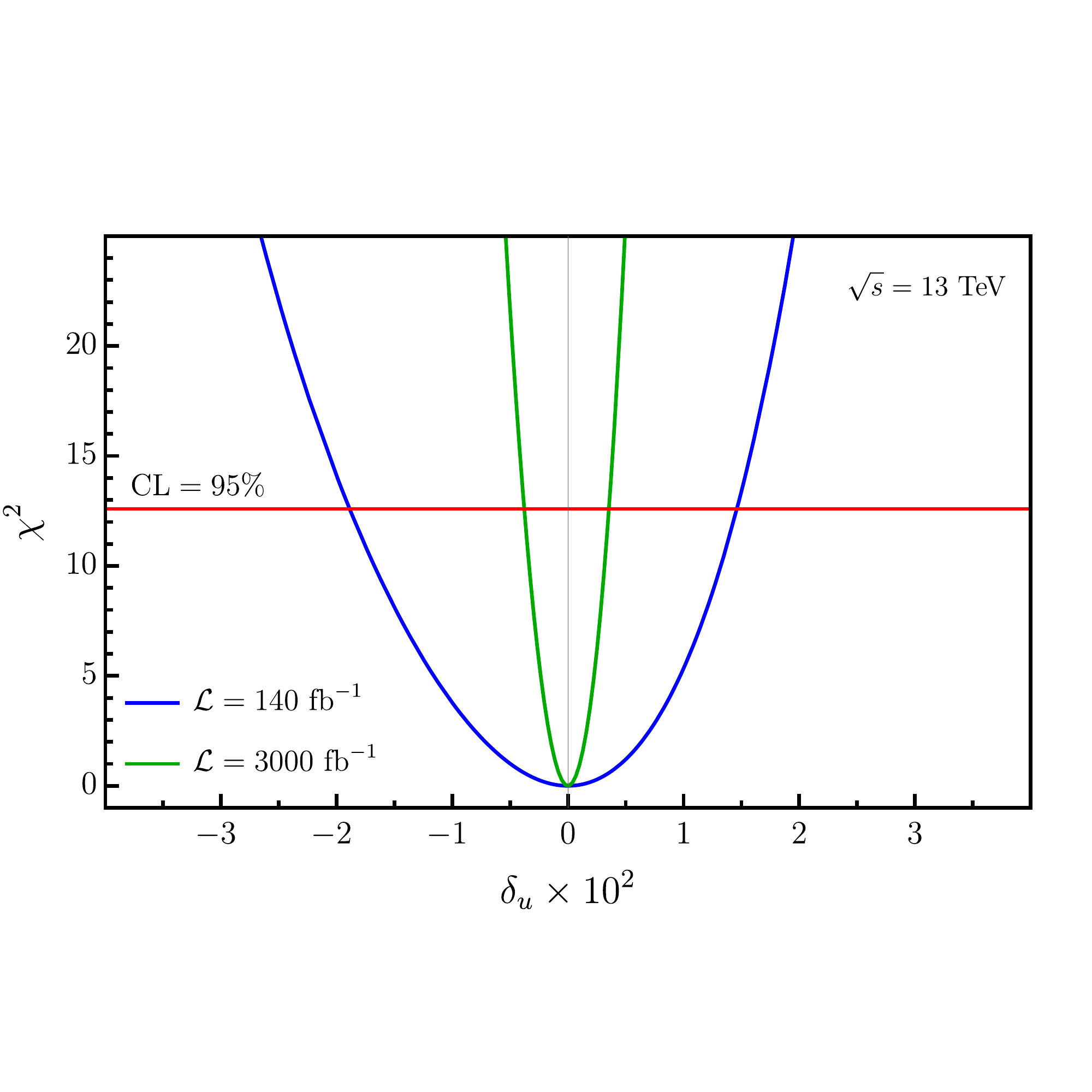}
  \includegraphics[width=.49\linewidth]{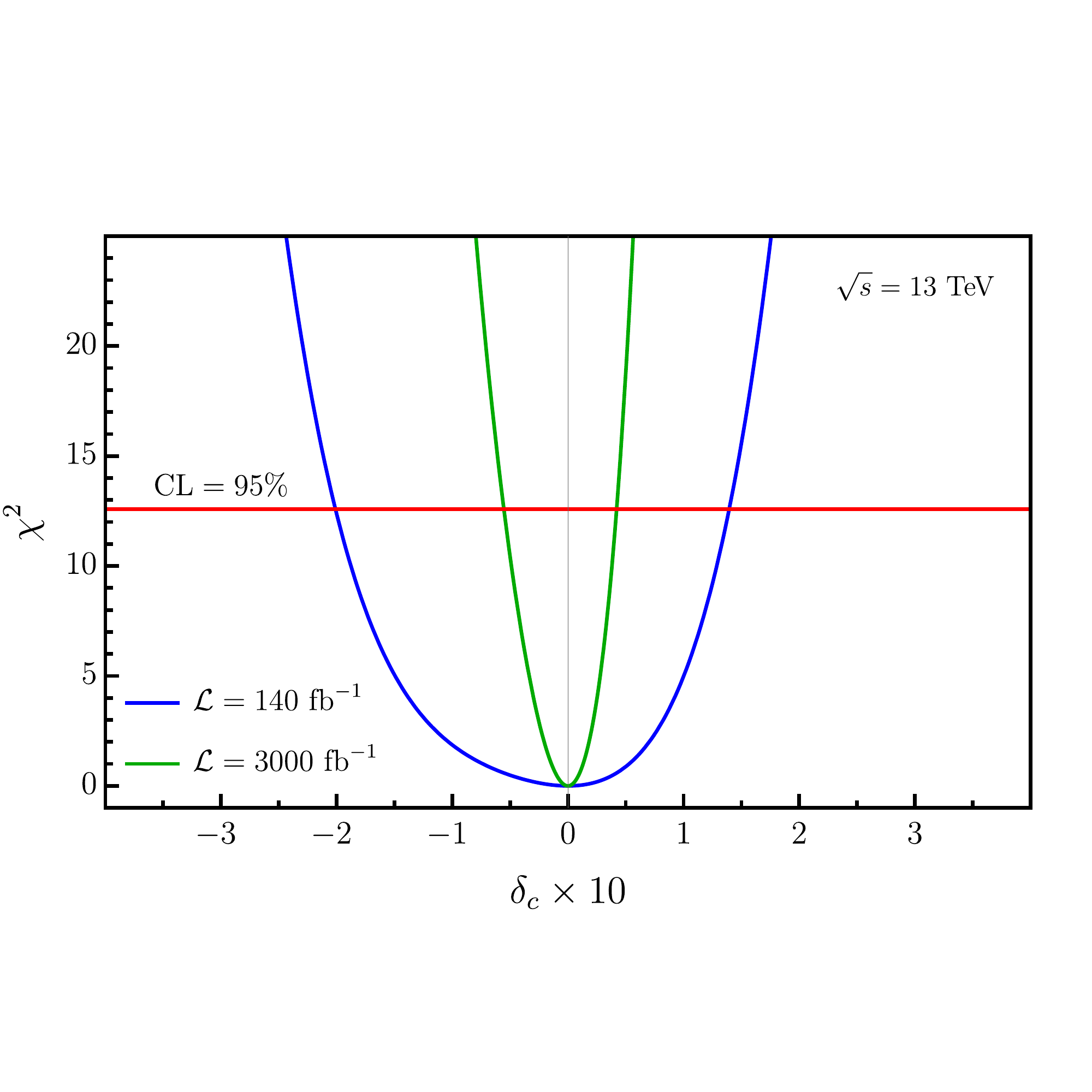}
  \\\vspace{-2.5cm}
  \includegraphics[width=.49\linewidth]{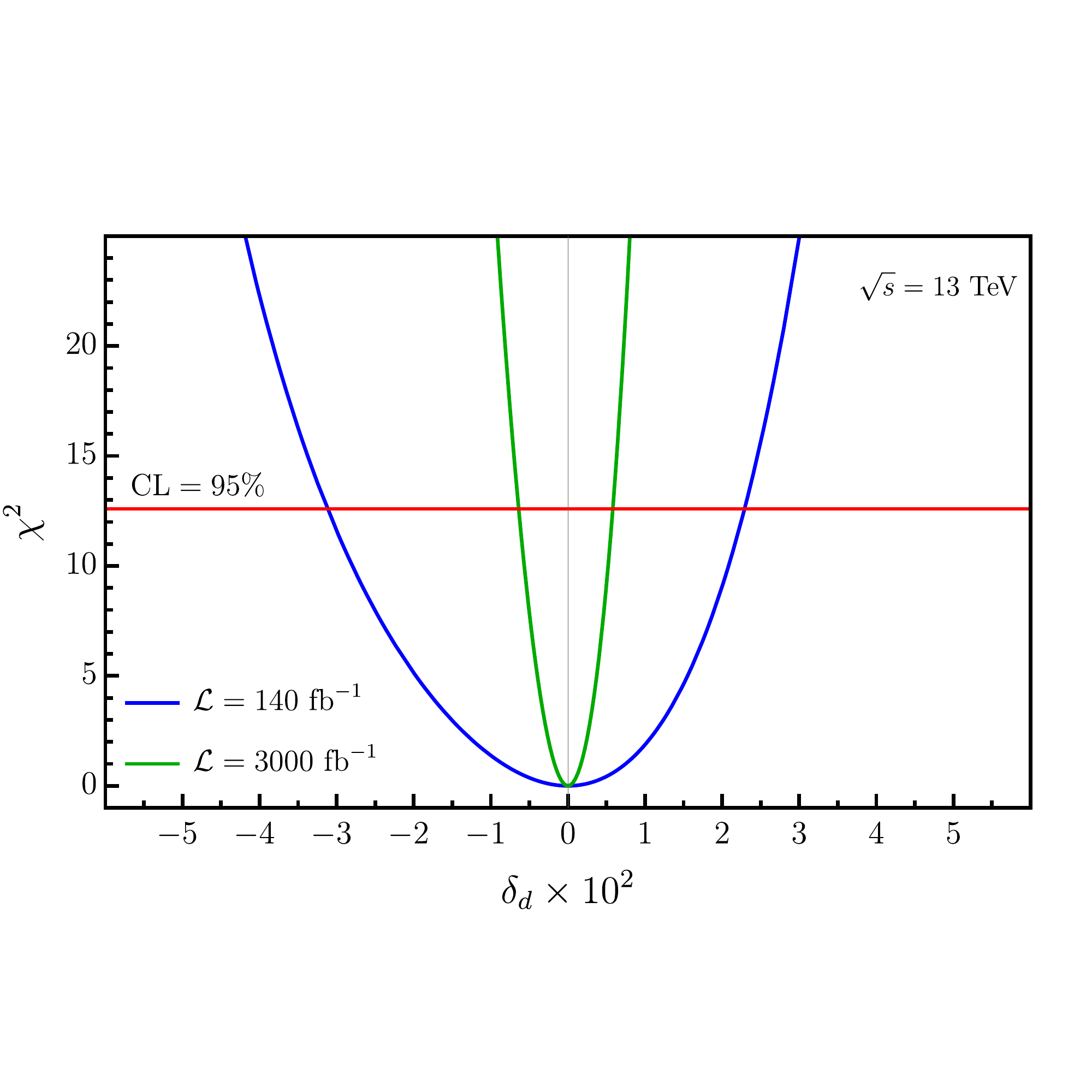}
  \includegraphics[width=.49\linewidth]{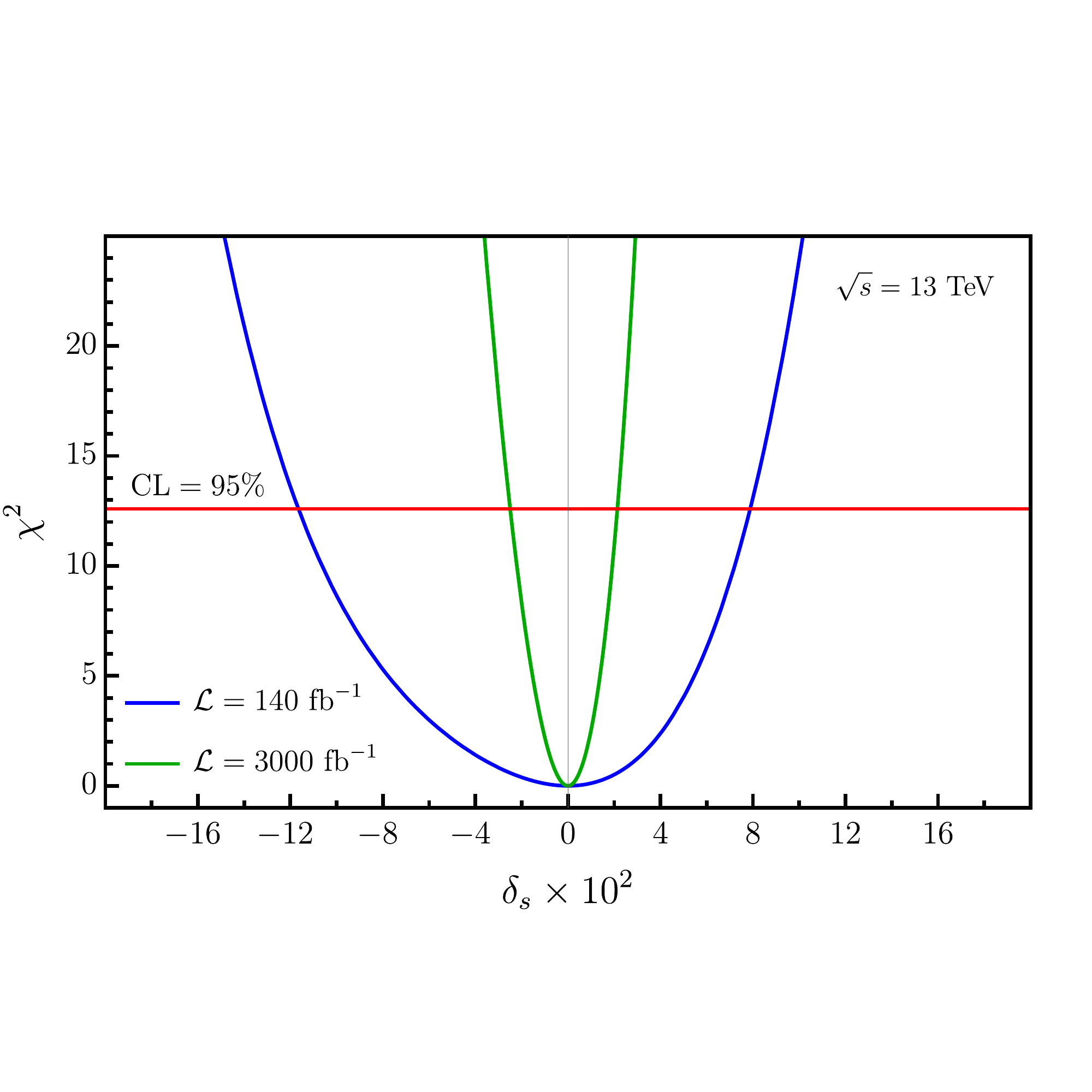}
  \vspace{-1.5cm}
  \caption{\small The $\chi^2$ as a function of $\delta_u$ (first panel), $\delta_c$ (second panel), $\delta_d$ (third panel) and $\delta_s$ (fourth panel). The blue curve uses the run2 LHC data with integrated luminosity ${\cal L}=140~{\rm fb}^{-1}$, while the green curve shows the corresponding HL-LHC reach for an integrated luminosity ${\cal L}=3000~{\rm fb}^{-1}$. Values of the parameters yielding a $\chi^2$ value above the red horizontal line are excluded at 95\% CL. 
\label{fig:UU-LHC} 
}
\end{center}
\end{figure}
\begin{table}[h!]
  \small
\bc
\begin{tabular}{ccc}
  95\% CL &\hskip1cm (run2) ${\color{oucrimsonred} {\cal L}=140\ \text{fb}^{-1}}$
  \hskip0.5cm &  \hskip0.5cm (Hi-Lumi) ${\color{oucrimsonred} {\cal L}=3000\ \text{fb}^{-1}}$  \hskip0.5cm \\[0.2cm]
\hline\\
$\boxed{\delta_u}$ &\hskip1cm  { $-1.9\times 10^{-2}<\delta_u<1.5\times 10^{-2}$}
\hskip0.4cm &\hskip0.4cm { $-3.8\times 10^{-3}<\delta_u<3.5\times 10^{-3}$} \hskip0.4cm \\[0.4cm] 
$\boxed{\delta_d}$ &\hskip1cm  { $-3.1\times 10^{-2}<\delta_d<2.3\times 10^{-2}$}
\hskip0.4cm &\hskip0.4cm { $-6.4\times 10^{-3}<\delta_d< 5.8\times 10^{-3}$} \hskip0.4cm \\[0.4cm]
$\boxed{\delta_s}$ &\hskip1cm  { $-1.2\times 10^{-1}<\delta_s<7.9\times 10^{-2}$}
\hskip0.4cm &\hskip0.4cm { $-2.5\times 10^{-2}<\delta_s<2.1\times 10^{-2}$} \hskip0.4cm \\[0.4cm]
$\boxed{\delta_c}$ &\hskip1cm  { $-2.0\times 10^{-1}<\delta_c<1.4\times 10^{-1}$}
\hskip0.4cm &\hskip0.4cm {$-5.5\times 10^{-2}<\delta_c<4.2\times 10^{-2}$} \hskip0.4cm \\[0.4cm]
$\boxed{\delta_b}$ &\hskip1cm  { $-3.5\times 10^{-1}<\delta_b<2.6\times10^{-1}$}
\hskip0.4cm &\hskip0.4cm { $-1.4\times 10^{-1}<\delta_b<8.8\times 10^{-2}$} \hskip0.4cm \\[0.4cm]
   \hline%
\end{tabular}
\caption{ \small\label{tab:delta-LHC} Limits at 95\% CL on the diagonal parameters $\delta_q$ obtained with LHC run2 data and corresponding HL-LHC projections.}
\ec
\end{table}

As mentioned before, in performing our analysis we retained only one non-vanishing $\delta_q$ contribution at a time, thereby neglecting also possible correlations between these parameters. This choice is meant to render the analysis as model-independent as possible although, within a given new physics model in which the $\delta_q$ terms are sourced by the parameters of the theory, correlation are expected. In this regard, the analytic expression we provided can be employed to straightforwardly perform the proposed test within any such framework accounting for the implied correlations between the $\delta_q$ factors.


\subsection{Future circular collider}

We extend the analysis to gauge the reach of the proposed method at the future
circular hadron collider FCC-hh~\cite{Mangano:2017tke,FCC:2018byv}, operating at $\sqrt{S}= 100~{\rm TeV}$. For the computation of the cross section in \eq{diffcross} we use again the {\tt PDF4LHC21} PDFs given in Ref.~\cite{PDF4LHCWorkingGroup:2022cjn} with $\sqrt{S}=100$ TeV and factorization scale $\mWW$. 

The distribution in $\mWW$, obtained by integrating the cross section on the full solid angle, is discretized over 9 benchmark bins defined by
\bea
\small
\Delta_i^{\rm FCC}&=&\Big\{
\left[2\MW/{\rm GeV},~500\right],~~\left[500,~1000\right],~~\left[1000,~1500\right],~~\left[1500,~2000\right],~~\left[2000,~3000\right],
\nonumber \\
&&~~\left[3000,~4000\right],~~\left[4000,~5000\right],~~\left[5000,~6000\right],~~\left[>6000 ~\right]
\Big\}\, ,
\label{FCCbins}
\eea
in units of GeV. In order to constrain the $\delta_q$ parameters we then introduce the following $\chi^2$ test
\be
\chi^2(\delta_q)=\sum_{i=1}^9
\frac{\Big(N_i(\delta_q)-N^{\rm SM}_{i}\Big)^2}
{N^{\rm SM}_{i}}\leq 16.919\quad(95\%\,{\rm CL})\, ,
\label{chi2FCC}
\ee
where $N_i(\delta_q)$ and $N^{\SM}_{i}= N_i(0)$ are the expected number of events in the $\Delta_i^{\rm FCC}$ bin in the presence of unitarity violation and in the SM, respectively.  We set the related uncertainties to $\sqrt{N^{\SM}_{i}}$ assuming that the experimental error will be dominated by the statistical one and include the NNLO QCD corrections accounting for a $K$-factor of $K\sim 1.58$, at $\sqrt{S}=100~{\rm TeV}$ ~\cite{deFlorian:2016uhr}, in the computation of the number of events. For the integrated luminosity we consider the benchmark scenarios of ${\cal L}=30~{\rm ab}^{-1}$~\cite{FCC:2018byv}.

\begin{table}[h!]
  \small
\bc
\begin{tabular}{cc}
  95\% CL &\hskip1cm (FCC-hh) ${\color{oucrimsonred} {\cal L}=30\ \text{ab}^{-1}}$
  \hskip0.5cm \\[0.2cm]
\hline\\
$\boxed{\delta_u}$ &\hskip1cm  { $-9.5\times 10^{-5}<\delta_u<9.2\times 10^{-5}$}
\hskip0.4cm \\[0.4cm]
$\boxed{\delta_d}$ &\hskip1cm  { $-1.3\times 10^{-4}<\delta_d<1.3\times 10^{-4}$ }
\hskip0.4cm \\[0.4cm]
$\boxed{\delta_s}$ &\hskip1cm  {$-2.4\times 10^{-4}<\delta_s<2.4\times 10^{-4}$ }
\hskip0.4cm
\\[0.4cm]
$\boxed{\delta_c}$ &\hskip1cm  {$-3.9\times 10^{-4}<\delta_c<3.9\times 10^{-4}$  }
\hskip0.4cm
\\[0.4cm]
$\boxed{\delta_b}$ &\hskip1cm  {$-7.4\times 10^{-4}<\delta_b<7.3\times 10^{-4}$  }
\hskip0.4cm
\\[0.4cm]
   \hline
\end{tabular}
\caption{ \small\label{tab:delta-FCC}Forecasts for the 95\% CL bounds obtained for the parameters $\delta_q$, $q=u,d,s,c,b$ at the FCC-hh collider with  $\sqrt{S}=100~{\rm TeV}$ and integrated luminosity of $30~{\rm ab}^{-1}$.}
\ec
\end{table}

In Tab.~\ref{tab:delta-FCC} we report the bounds obtained at 95\% CL for the  $\delta_q$ parameters. As we can see, the $W^+W^-$ production at 100 TeV can potentially test the unitarity of the CKM matrix at a level of precision well on par with the present sensitivities obtained in flavor physics experiments \eqs{CKMuu}{CKMcc}.
In particular, the test restricts the $\delta_{u,d}$ and $\delta_{s}$ parameters below $\sim 1\times 10^{-4}$ and $\sim 3\times 10^{-4}$, respectively, while the magnitude of $\delta_c$ and $\delta_b$ can be probed to corresponding precisions of $\sim 4\times10^{-4}$ and $\sim 7\times10^{-4}$. Clearly, the inclusion of the background, detector effects and potential systematic effects would deteriorate the constraints show in Tab.~\ref{tab:delta-FCC}, to an extent that can only be quantified after the machine will be well-understood. 

\section{Summary}
\label{sec:Results} 
We explore the potential of the $W^+W^-$ production cross section at hadron colliders to test violations of unitarity in the quark mixing, parameterized by the CKM matrix. By using the most recent ATLAS measurements of the observable, the proposed test constrains the diagonal unitarity violation parameters defined in \eq{xiq} to a level of $|\delta_{u,d}|\lsim {\cal O}(10^{-2})$ and $|\delta_{c,s}|\lsim {\cal O}(10^{-1})$ for first and second quark generation, respectively. The amount of unitarity violation admitted by the LHC run2 data in the $b$ sector is instead $|\delta_b|\lsim 30\%$. We estimate that the bounds could improve roughly by a factor of order $\sim$ 4--5 and $\sim$ 3--4 for $\delta_{u,d,s}$ and $\delta_{c,b}$, respectively, with the planned high-luminosity LHC phase.
We repeat the analysis to gauge the potential of the future FCC-hh hadron collider, operating at $\sqrt{S}=100~{\rm TeV}$ and with a benchmark total integrated luminosity of $30~{\rm ab}^{-1}$. Due to the fact that the sensitivity of the $pp\to W^+W^-$ cross section to the $\delta_q$ parameters grows quadratically with the $W^+W^-$ invariant mass, the constraints posed by the proposed unitarity test largely improve. Depending on the experimental sensitivity that will be achieved for the $W^+W^-$ cross section in the range of high $W^+W^-$ invariant mass, $\mWW \gsim 1~{\rm TeV}$, the method can produce limits of order $|\delta_{q}|\lsim {\cal O}(10^{-4})$  comparable, or even stronger, than the present bounds from flavor physics. The investigation can be straightforwardly extended to test also off-diagonal contributions to unitarity violations, corresponding to products of different rows, or columns, of the CKM matrix. Although it is currently difficult to assess the full potential of the proposed method, our work shows that it can provide independent strong bounds apt to complement the results of more traditional tests of unitarity violation.

\section*{Acknowledgements}
The authors thank M. Fabbrichesi, M. Grazzini, T. Lagouri, R. Mazini and H. Veerm\"ae for useful discussions. E.G. acknowledges the Department of Theoretical Physics of CERN  for the kind hospitality during the preparation of this work. This work was supported by the Estonian Research Council grants PRG803, RVTT3 and by the CoE program grant TK202 ``Fundamental Universe'’.

\begin{appendices}
\section{Analytical expressions
  \label{sec:Appendix}}
We provide below the complete analytical expressions for the quantities 
$\Delta \xbar{{\cal M}}^{\; q\bar q}_{1,2}$ appearing in \eq{M2tot}. For the up-type quarks, $U=u,c$, we have

\bea 
\Delta \xbar{{\cal M}}^{\; U\bar U}_{1}&=&
\frac{4\alpha_{\W}^2\pi^2}{3 N_c \left(1-\betaW^2\right)^2\left(1+2\betaW\Ct+\betaW^2\right)^2\left(\MZ^2-s\right)}
\Bigg\{3 s \left(1 - \betaW^2\right) \Big[-8 - 24 \betaW \Ct 
\nonumber\\
&-& \betaW^2 \left(7 + 9 \Ct^2\right) + 16 \betaW^3 \Ct
+ 2 \betaW^4 \left(7 + 3 \Ct^2 + 2 \Ct^4\right)
+ 8 \betaW^5 \Ct^3 -  3 \betaW^6 \left(1 - \Ct^2\right)
\Big] \nonumber\\
&+& 2 \MZ^2 \Big[12 + 16 \betaW \Ct \left(3 -\ssW^2\right)+ \betaW^2 \Big(27 + \Ct^2 \left(33 - 26 \ssW^2\right) - 38 \ssW^2\Big)
  \nonumber\\
&+& 4 \betaW^3 \Ct \Big(3 - 24 \ssW^2 - \Ct^2 \left(3 - 4 \ssW^2\right)\Big)
  - 2 \betaW^4 \Big(15 + \ssW^2 + 3 \Ct^2 \left(3 + \ssW^2\right) + \Ct^4 \left(6 - 4 \ssW^2\right)\Big)
\nonumber\\
&-& 4 \betaW^5 \Ct \left(9 + 3 \Ct^2 - 20 \ssW^2\right) 
+ \betaW^6 \Big(3 + 30 \ssW^2 + 8 \Ct^4 \ssW^2 - \Ct^2 \left(15 - 26 \ssW^2\right)\Big)\nonumber\\
&+& 16 \betaW^7 \Ct^3 \ssW^2 - 6 \betaW^8 \left(1 - \Ct^2\right) \ssW^2
\Big]
\Bigg\}\,,
\label{eq:A1}
\eea

\bea
\Delta \xbar{{\cal M}}^{\; U\bar U}_{2}&=&
\frac{4\alpha_{\W}^2\pi^2}{N_c   \left(1-\betaW^2\right)^2\left(1+2\betaW\Ct+\betaW^2\right)^2}
\Bigg\{4 + 16 \betaW \Ct + \betaW^2 (9 + 11 \Ct^2) + 4 \betaW^3 \Ct (1 - \Ct^2)
\nonumber\\
&-&  2 \betaW^4 (5 + 3 \Ct^2 + 2 \Ct^4)
- 4 \betaW^5 \Ct (3 + \Ct^2) + \betaW^6 (1 - 5 \Ct^2)
\Bigg\}\,.
\label{eq:A2} 
\eea

The analytical forms of the $\Delta \xbar{{\cal M}}^{\; D\bar D}_{1,2}$ terms involving down-type quarks, $D=d,s,b$, can be obtained from the ones above through the global substitution $\Ct\to -\Ct$ and by the replacement $\ssW^2\to \frac{1}{2}\ssW^2$ in the terms appearing inside the curly brackets. Definitions for the symbols appearing in \eqs{eq:A1}{eq:A2} can be found in section \ref{sec:DY}.

\end{appendices} 
\newpage
\bibliographystyle{jhep}
\bibliography{WW-PLB-rev.bib}


\end{document}